\documentclass[aps,prb,twocolumn,floats]{revtex4-1}
\usepackage{color,graphicx,pstricks}
\usepackage{hyperref}
\hypersetup{colorlinks=true}
\usepackage{etoolbox}
\apptocmd{\sloppy}{\hbadness 10000\relax}{}{}

\begin{document}  

\title{Nanoclustering phase competition induces the resistivity hump in                                                 
colossal magnetoresistive manganites}
%\author{Kalpataru Pradhan$^{1,2}$ and Seiji Yunoki$^{2,3,4}$}
\author{Kalpataru Pradhan$^{1,2}$}
\email[email: ]{kalpataru.pradhan@saha.ac.in}
\author{Seiji Yunoki$^{2,3,4}$}
\email[email: ]{yunoki@riken.jp}
\affiliation{$^1$CMP Division, Saha Institute of Nuclear Physics, 
HBNI, Kolkata 700064, India\\
$^2$Computational Quantum Matter Research Team, RIKEN, 
Center for Emergent Matter Science (CEMS), Saitama 351-0198, Japan\\
$^3$Computational Condensed Matter Physics Laboratory, 
RIKEN, Wako, Saitama 351-0198, Japan\\
$^4$Computational Materials Science Research Team, RIKEN Advanced 
Institute for Computational Science (AICS), Hyogo 650-0047, Japan}

\date{\today}

\begin{abstract}
Using a two-band double-exchange model with Jahn-Teller lattice distortions and 
super-exchange interactions, supplemented by quenched disorder, at electron density 
$n=0.65$, we explicitly demonstrate the coexistence of the $n$ = 1/2-type ($\pi, \pi$) 
charge-ordered and the ferromagnetic nanoclusters above the ferromagnetic transition 
temperature $T_{\rm c}$ in colossal magnetoresistive (CMR) manganites. The resistivity 
increases due to the enhancement of the volume fraction of the charge-ordered and the 
ferromagnetic nanoclusters with decreasing the temperature down to $T_{\rm c}$. The 
ferromagnetic nanoclusters start to grow and merge, and the volume fraction of the 
charge-ordered nanoclusters decreases below $T_{\rm c}$, leading to the sharp drop in 
the resistivity. By applying a small external magnetic field $h$, we show that the 
resistivity above  $T_{\rm c}$ increases, as compared with the case when $h=0$, a fact 
which further confirms the coexistence of the charge-ordered and the ferromagnetic 
nanoclusters. In addition, we show that the volume fraction of the charge-ordered 
nanoclusters decreases with increasing the bandwidth and consequently the resistivity 
hump diminishes for large bandwidth manganites, in good qualitative agreement with 
experiments. The obtained insights from our calculations provide a complete pathway 
to understand the phase competition in CMR manganites. 
\end{abstract}

\maketitle

%%%%%%%%%%%%%%%%%%%%%%%%%%%%%%%%%%%%%%%%%%%%%%%%%%%%%%%%%%%%%%%%%%%%%%%%

%\section*{Introduction}
\section{Introduction}

The strong coupling between charge, spin, orbital, and lattice degrees of freedom, 
supplemented by weak disorder, leads to unusual colossal response phenomena in 
manganites~\cite{tok-book,dag-book,dagotto,tokura}. There has been intense focus over 
last two decades to validate the phase coexistence/competition 
scenario~\cite{uehara,zhang,faeth,sarma,zuo,koo,budhani,murakami,tao,burkhardt,horibe,bingham,rawat,tao1}, 
which is believed to be a necessary ingredient to explain the resistivity hump in 
colossal magnetoresistive (CMR) manganites. Furthermore, understanding the phase 
competition in bulk manganites helps in designing low-dimensional manganite 
nanostructures with emergent physical phenomena~\cite{liang,shaoa,zhang1}.

\begin{figure} [t]
\centerline{
\includegraphics[width=8.40cm,height=7.2cm,clip=true]{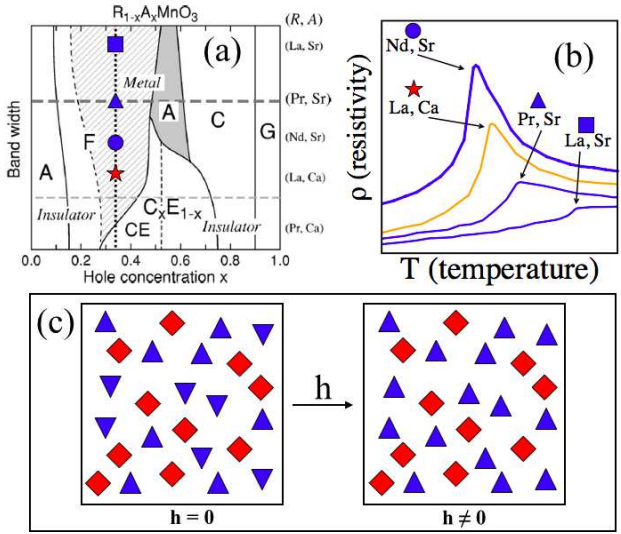}}
\caption{(Color online) (a) Schematic low-temperature phase diagram of manganites 
R$_{1-x}$A$_x$MnO$_3$ for different R (rare-earth) and A (alkaline-earth) elements 
(adopted from Ref.~\citenum{kajimoto}). F, A and CE-type phases near $x = 1-n = 0.35$ are of 
interests to the present study and they  denote ferromagnetic (FM), A-type antiferromagnetic, 
and CE-type antiferromagnetic phases, respectively. (b) Schematic to illustrate the 
temperature dependence of the resistivity for different bandwidth manganites at 
$x = 1-n\sim1/3$. Same symbols in (a) and (b) are used to indicate the combination of R and A. 
(c) Schematic to show the phase coexistence between charge-ordered (red diamonds) and 
ferromagnetic (blue triangles) nanoclusters without and with a small external magnetic field 
$h$ above the FM $T_{\rm c}$. The direction of a triangle implies the overall spin 
orientation in each FM nanoclusters (see the text for details). 
}\label{fig:schematic}
\end{figure}

The manganites, materials of the form R$_{1-x}$A$_{x}$MnO$_3$ where R is rare-earth 
(La, Nd, Pr, etc.) and A is alkaline-earth (Ca, Sr, and Ba) elements, are best known 
for their CMR property in the doping range $x = 1-n$ = 0.3--0.4 ($n$: electron density). 
The bandwidth of the manganites increases with the mean radius $r_{\rm A}$ of R and A 
ions~\cite{kajimoto,martin}, and the size mismatch between the two radii controls the 
strength of the cation disorder~\cite{attfield,hwang}. Figure~\ref{fig:schematic}(a), 
reproduced from Ref.~\citenum{kajimoto}, schematically summarizes a phase diagram of 
the manganites. Low temperature magnetic states for $n$ = 0.5 and 0.65 are also listed in 
Table~\ref{table} along with ionic radii of different R and A ions.

The phase competition arising due to the proximity of variety of phases, in presence of 
cationic disorder, often leads to novel phenomena such as inhomogeneities, phase 
coexistence, and percolative transport~\cite{dagotto,tokura}. Recent 
experiments~\cite{burkhardt,horibe,bingham,rawat} show that nanoscale ferromagnetic (FM) 
regions and $n$ =1/2-type ($\pi, \pi$) charge-ordered (CO)~\cite{uehara} regions, which 
characterizes the CE-type phase, coexist above the FM transition temperature $T_{\rm c}$. 
The precise microscopic origin of the phase competition above $T_{\rm c}$ is still elusive 
theoretically~\cite{ps-q,open-q}. The absence of FM correlations above $T_{\rm c}$ fails 
to establish the phase competition scenario in CMR manganites in recent theoretical 
studies~\cite{sen2,sen3}.

%%%%%%%%%%%%%%%%%%%%%%%%%%%%%%%%%%%%%%%%%%%%%%%%%%%%%%%%%%%%%%%%%%%%%%%%
%%%%%%%%%%%%%%%%%%%%%%%%%%%%%%%%%%%%%%%%%%%%%%%%%%%%%%%%%%%%%%%%%%%%%%%%

In this paper, we explicitly demonstrate the phase competition between the $n$ =1/2-type 
($\pi, \pi$) CO nanoclusters and the FM nanoclusters at $n$ = 0.65, relevant to the CMR 
manganites, above the FM $T_{\rm c}$. Our study 
reveals that this phase competition induces the resistivity hump, and is expected only 
in those manganites for which the ground state is the CE-type phase at $n\sim0.5$. With 
increasing the bandwidth, the volume fraction of the CO nanoclusters decreases and 
consequently the resistivity hump around $T_{\rm c}$ vanishes. 

\section{Model Hamiltonian and Method}

We consider the following two-band model Hamiltonian~\cite{dagotto}, 
a minimal model that is required to capture the essential phases in 
manganites~\cite{yunoki,sanjeev,pradhan1,pradhan2}, in two spatial dimensions:
\begin{eqnarray}
H &=& - \sum_{\langle ij \rangle, s}\sum_{\alpha,\beta}
t_{\alpha \beta}^{ij}
 c^{\dagger}_{i \alpha s} c^{~}_{j \beta s}
 - J_H\sum_i {\bf S}_i\cdot{\mbox {\boldmath $\sigma$}}_i 
+ J\sum_{\langle ij \rangle} {\bf S}_i\cdot{\bf S}_j \cr
&& - \lambda \sum_i {\bf Q}_i\cdot{\mbox {\boldmath $\tau$}}_i
+ {K \over 2} \sum_i {\bf Q}_i^2 +
{\sum_i(\epsilon_i - \mu) n_i}, 
\end{eqnarray}
\noindent
where $c_{i\alpha s}^\dag$ is the electron creation operator at site $i$ with orbital 
$\alpha\,=(a, b)$ and spin $s\,(=\uparrow,\downarrow)$. The kinetic energy term includes 
inter- and intra-orbital nearest-neighbor hopping $t^{ij}_{\alpha \beta}$, i.e. 
$t_{a a}^x= t_{a a}^y \equiv t$, 
$t_{b b}^x = t_{b b}^y\equiv t/3 $, $t_{a b}^x = t_{b a}^x \equiv -t/\sqrt{3} $, 
and $t_{a b}^y = t_{b a}^y \equiv t/\sqrt{3} $ along $x$ and $y$ directions, 
where $a$ and $b$ refer to two Mn $e_g$ orbitals $d_{x^2-y^2}$ and $d_{3z^2-r^2}$, 
respectively. Hund's coupling $J_H$ is between $t_{2g}$ spin ${\bf S}_i$ and $e_g$ 
electron spin {\boldmath $\sigma$}$_i$ at site $i$. Here, we adopt the double-exchange 
limit, i.e. $J_H \rightarrow \infty$, as $t$ ($\sim$0.2--0.5 eV) 
is estimated to be much smaller than $J_H$ ($\sim$2 eV)~\cite{dagotto}. 
$J$ is the antiferromagnetic super-exchange between nearest neighboring 
$t_{2g}$ spins. $\lambda$ represents the electron-phonon coupling between Jahn-Teller 
phonons ${\bf Q}_i$ and $e_g$ electrons in the adiabatic limit. We treat all 
${\bf S}_i$ and ${\bf Q}_i$ as classical variables~\cite{yunoki1,class-ref,yunoki2}, and 
the stiffness $K$ of Jahn-Teller modes and $|{\bf S}_i|$ are set to be 1. 
We also choose a typical value of $J/t$ = 0.1~\cite{yunoki}. 
Effects of disorder is taken into account by the $\sum_i \epsilon_i n_i$ term 
where $\epsilon_i$ is the quenched binary disorder potential with values 
$\pm \Delta$. $\mu$ is the chemical potential. To analyze the effect of external 
magnetic field ${\bf h}=h\hat{\bf z}$, wherever necessary, we add a Zeeman coupling term 
$-\sum_i {\bf h}\cdot{\bf S}_i$ to the Hamiltonian. We measure $\lambda$, $\Delta$, $h$ 
and temperature $T$ in units of $t$.

%%%%%%%%%%%%%%%%%%%%%%%%%%%%%%%%%%%%%%%%%%%%%%%%%%%%%%%%%%%%%%%%%%%%%%%%
\begin{table}[t]
\centering
\begin{center}
%\begin{tabular}{|l|l|l|l|l|l|l|l|}
\begin{tabular}{|c|c||c|c||c|c|}
\hline
\multicolumn{4}{|c||}{experimentally observed magnetic phases} &\multicolumn{2}{c|}{parameters}   \\ \hline 
 {manganites} & radii (R,A) in \AA & n = 0.65 & n = 0.5  & $\lambda$  &  $\Delta$ \\\hline

 PCMO & 1.29, 1.34  &   CE-type & CE-type &      $1.75$ & $0.1$  \\ \hline
 LCMO & 1.36, 1.34  &   FM-M    & CE-type &      $1.70$ & $0.05$ \\ \hline
 NSMO & 1.27, 1.44  &   FM-M    & CE-type &      $1.65$ & $0.2$  \\ \hline
 PSMO & 1.29, 1.44  &   FM-M    & A-type  &      $1.55$ & $0.2$ \\ \hline
 LSMO & 1.36, 1.44  &   FM-M    & FM-M    &      $1.50$ & $0.1$  \\ \hline
\end{tabular}
\caption{Magnetic states observed experimentally at low temperatures 
[from Fig.~\ref{fig:schematic}(a)] for Pr$_{1-x}$Ca$_{x}$MnO$_3$ (PCMO), 
La$_{1-x}$Ca$_{x}$MnO$_3$ (LCMO), Nd$_{1-x}$Sr$_{x}$MnO$_3$ (NSMO), 
Pr$_{1-x}$Sr$_{x}$MnO$_3$ (PSMO), and La$_{1-x}$Sr$_{x}$MnO$_3$ (LSMO) at $n = 1-x = 0.65$ 
and $0.5$. FM-M stands for FM metal. 
$\lambda$ and $\Delta$ values used in our calculations to qualitatively 
reproduce the experimental results and ionic radii of R and A elements~\cite{tokura} 
are also given. 
}
\label{table}
\end{center}
\end{table}
%%%%%%%%%%%%%%%%%%%%%%%%%%%%%%%%%%%%%%%%%%%%%%%%%%%%%%%%%%%%%%%%%%%%%%%%

An exact diagonalization scheme is applied to the mobile $e_g$ electrons in the 
background of classical spins ${\bf S}_i$ and lattice distortions ${\bf Q}_i$, and a 
Monte Carlo method is employed for classical variables. This spin-fermion Monte Carlo 
method adequately takes spatial fluctuations into account that is necessary to study 
the inhomogeneities. In order to handle large system sizes (up to $N = 24 \times 24$ 
sites), we employ a Monte Carlo technique based on traveling cluster 
approximation~\cite{tca-ref,sanjeev,pradhan1}. The randomized classical spins 
${\bf S}_i$ and lattice distortions ${\bf Q}_i$ are annealed starting from a high 
temperature in an arbitrary quenched disorder configuration. All physical quantities 
such as magnetization and resistivity are averaged over ten such different disorder 
configurations in addition to the thermal averages taken during the Monte Carlo 
simulations.

%\section*{Results and Discussion}

\section{Phase Diagram ($\lambda = 1.7$ and $J = 0.1$)}

\begin{figure} [t]
\centerline{
\includegraphics[width=8.75cm,height=7.25cm,clip=true]{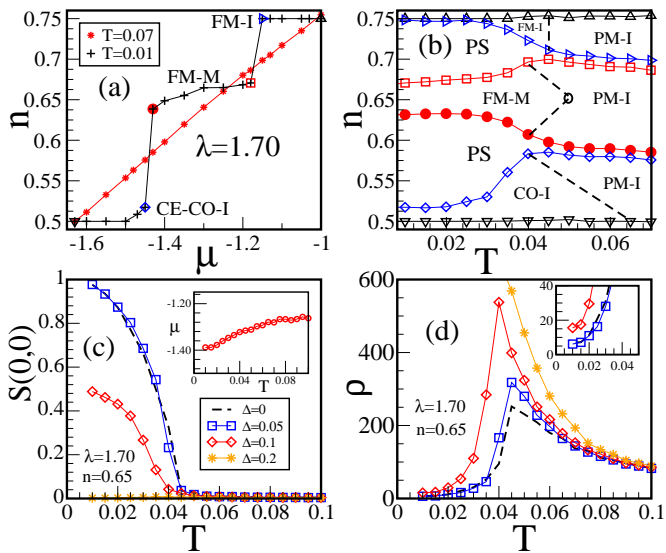}}
\caption{(Color online) 
(a) $n$ dependence on $\mu$ for two different temperatures $T = 0.01$ and 0.07. 
(b) The $n$-$T$ phase diagram, containing the CO and FM phases. 
PS, I, and M denote phase separation, insulator, and metal, respectively. 
The corresponding phase boundaries at $T=0.01$ are indicated in (a) using the same 
symbols as in (b). $\Delta=0$ in (a) and (b). $T$ dependence of (c) the FM structure 
factor $S(0,0)$ and (d) the resistivity $\rho$ in units of ${\hbar a}/{\pi  e^2 }$ at 
$n$ = 0.65 for different $\Delta$ values indicated in (c). The inset in (c) shows the 
$\mu$ vs. $T$, required to set the desired $n = 0.65$. The inset in (d) shows the 
enlarged plot at low temperatures. $\lambda=1.7$ for all figures. 
}\label{fig:pd}
\end{figure}

Figures~\ref{fig:pd}(a) and \ref{fig:pd}(b) show that the ground state is FM metallic 
in $e_g$ electron density $n$ = 0.63--0.67 for $\lambda = 1.7$, and is separated from 
the CE-type insulating phase at $n = 0.5$ and the FM insulating phase at 
$n = 0.75$~\cite{pradhan1} by phase separation windows at $T = 0.01$. Indeed, $n$ jumps 
from 0.52 to 0.63 and from 0.67 to 0.75 in the $n$ vs. $\mu$ curve. However, for high 
temperatures, $n$ monotonically increases with $\mu$ without any discontinuities 
[see Fig.~\ref{fig:pd}(a) for $T = 0.07$]. Figure~\ref{fig:pd}(b) summarizes the $n$--$T$ 
phase diagram in the electron density range $n$ = 0.5--0.75, showing different phases. 

%%%%%%%%%%%%%%%%%%%%%%%%%%%%%%%%%%%%%%%%%%%%%%%%%%%%%%%%%%%%%%%%%%%%%%%%
%%%%%%%%%%%%%%%%%%%%%%%%%%%%%%%%%%%%%%%%%%%%%%%%%%%%%%%%%%%%%%%%%%%%%%%%

%%%%%%%%%%%%%%%%%%%%%%%%%%%%%%%%%%%%%%%%%%%%%%%%%%%%%%%%%%%%%%%%%%%%%%%%
\begin{figure} [t]
\centerline{
\includegraphics[width=8.75cm,height=7.25cm,clip=true]{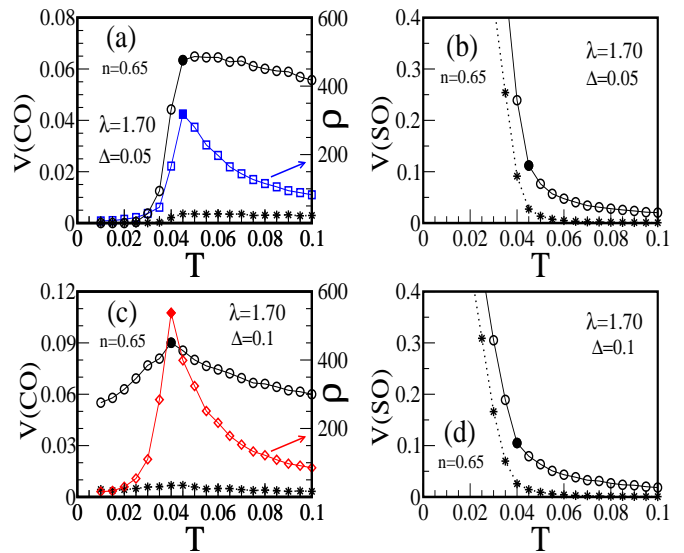}}
\caption{(Color online) 
$T$ dependence of the volume fraction of the 
CO nanoclusters V(CO) and resistivity $\rho$ (in units of 
${\hbar a}/{\pi  e^2 }$) for (a) $\Delta$ = 0.05 and (c) $\Delta$ = 0.1. The volume 
fraction of FM nanoclusters V(SO) vs. $T$ for (b) $\Delta$ = 0.05 and (d) $\Delta$ = 0.1. 
Filled symbols indicate $T_{\rm c}$. 
$n=0.65$ and $\lambda=1.7$ for all figures. 
See text for dotted lines with star symbols. 
}\label{fig:vco}
\end{figure}
%%%%%%%%%%%%%%%%%%%%%%%%%%%%%%%%%%%%%%%%%%%%%%%%%%%%%%%%%%%%%%%%%%%%%%%%

We focus on $n$ = 0.65 that is relevant to the CMR manganites. For this purpose, 
$\mu$ is varied to fix $n$ for different temperatures [inset of Fig.~\ref{fig:pd}(c)]. 
The temperature dependence of the magnetic structure factor 
$S(\textbf{q})$ = ${1 \over N^2}$ $\sum_{ij}$
$\bf {\bf S}_i\cdot {\bf S}_j$ e$^{i\bf{q} \cdot ({\bf r}_i-{\bf r}_j)}$ at wave vector 
${\bf q} = (0, 0)$ (FM correlations) is shown in Fig.~\ref{fig:pd}(c) for different 
$\Delta$ values. The FM $T_{\rm c}$ remains the same for $\Delta$ = 0.05 as compared 
to $\Delta$ = 0 and decreases for $\Delta$ = 0.1. The corresponding resistivity 
in units of ${\hbar a}/{\pi  e^2 }$ ($a$: lattice constant) with temperature obtained 
by calculating the {\it dc} limit of the conductivity using the 
Kubo-Greenwood formalism~\cite{mahan-book,cond-ref} is shown in Fig.~\ref{fig:pd}(d). 
The magnitude of the resistivity hump around $T_{\rm c}$ increases with $\Delta$ and 
its position shifts to the lower temperature. However, the system remains insulating 
for $\Delta$ = 0.2 at all temperatures without any long range FM order. The intimate 
correlation between the onset of the ferromagnetism and the resistivity hump 
indicates that the metallic and the insulating phases compete each other near 
$T_{\rm c}$.

%\section*{Phase Competition between Charge-ordered and Ferromagnetic nanoclusters above $T_{\rm c}$}
\section{Phase Competition above $T_{\rm c}$}
 
In order to examine the phase competition, a measure of the volume fraction of the CO (or FM) 
regions is necessary because the structure factors in the momentum space are inefficient to 
probe the local ordering in the real space. Here we calculate the volume fraction of the 
$n$ = 1/2-type ($\pi, \pi$) CO nanoclusters V(CO) from the real space charge distribution. 
V(CO) is calculated by counting the fraction of sites for which the local density $n_i$ at 
site $i$, satisfying $n_i-0.5\ge$ 0.1, is surrounded by the four nearest neighbor sites $j$ 
with $0.5-n_j\ge$ 0.1 or vice versa. Similarly, the volume fraction of the FM nanoclusters, 
V(SO), is obtained by calculating the fraction of sites, say $i$, for which all 
${\bf S}_{\rm i}\cdot{\bf S}{\rm _j}$ $\ge$ 0.5 with the four nearest neighboring sites $j$.

As shown in Figs.~\ref{fig:vco}(a) and \ref{fig:vco}(c), V(CO) for $\Delta$ = 0.05 and 0.1, respectively, 
increases with decreasing $T$ until $T_{\rm c}$ and decreases thereafter. The volume fraction 
of the CO nanoclusters including the next nearest 
neighboring sites, plotted by dotted lines with star symbols in Figs.~\ref{fig:vco}(a) and 
\ref{fig:vco}(c), remains very small. Note that V(CO) decreases to zero for 
$\Delta$ = 0.05 but remains finite for $\Delta$ = 0.1 at low temperatures. The reminiscent 
of V(CO) for $\Delta$ = 0.1 correlates with the fact that the system is not fully FM even at 
$T$ = 0.01 where $S(0,0) \sim $ 0.5 [see Fig.~\ref{fig:pd}(c)]. This shows the fact that the 
CO nanoclusters coexist with the FM regions for $\Delta$ = 0.1 at low temperatures and as a 
result the resistivity is larger than in the clean systems, as shown in Fig.~\ref{fig:pd} (d).

It is apparently clear from Figs.~\ref{fig:vco}(a) and \ref{fig:vco}(c) that the resistivity 
increases with V(CO) until $T_{\rm c}$ and decreases below it. However, the enhancement of 
resistivity is steeper than V(CO), which indicates that electrons moving across the system are 
not only scattered from the CO nanoclusters but also from the FM nanoclusters that are possibly 
present in the system. Therefore, the logical next step is to reveal the presence of FM 
nanoclusters above $T_{\rm c}$. For this purpose, we plot V(SO) in Figs.~\ref{fig:vco}(b) and 
\ref{fig:vco}(d). With decreasing $T$, V(SO) increases up to $\sim0.1$ at $T_{\rm c}$ and then 
starts to grow in size. Eventually, the FM nanoclusters merge with each other at low temperatures. 
This is concluded from the fact that the volume fraction of the FM nanoclusters including the 
next nearest neighbor sites [dotted line with star symbols in Fig.~\ref{fig:vco}(b) and 
\ref{fig:vco}(d)] remains very small above $T_{\rm c}$, but increases below it for both 
$\Delta$ = 0.05 and 0.1. Typically, the strength of disorder is quantified by the variance of 
the ionic radii of R and A ions~\cite{attfield,hwang}. For example, cation mismatch is very 
small for La$_{1-x}$Ca$_{x}$MnO$_3$ (LCMO). In addition, the ground state of LCMO (at $x \sim 1/3$) 
is homogeneous~\cite{dag-book}. Therefore, we expect that $\Delta$ = 0.05 is more appropriate 
for LCMO (see TABLE~\ref{table}), and our calculations show FM nanoclusters coexisting with 
CO nanoclusters above $T_{\rm c}$ and changes to uniform FM metal at low temperatures.

%%%%%%%%%%%%%%%%%%%%%%%%%%%%%%%%%%%%%%%%%%%%%%%%%%%%%%%%%%%%%%%%%%%%%%%%
\begin{figure} [t]
\centerline{
\includegraphics[width=8.75cm,height=7.25cm,clip=true]{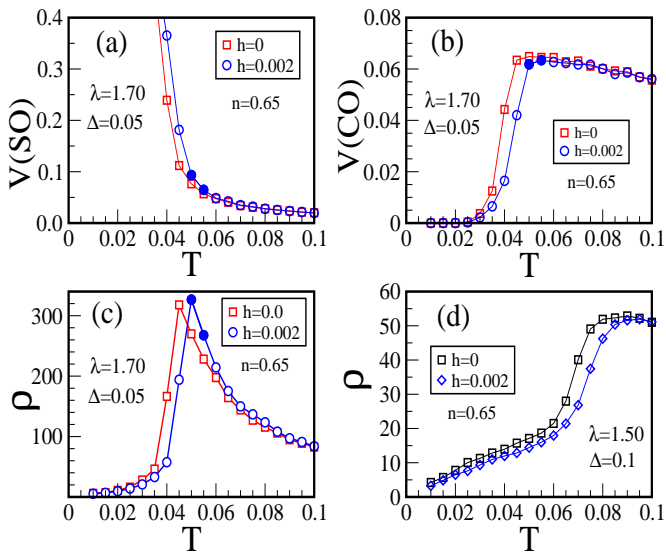}}
\caption{(Color online) 
$T$ dependence of the volume fraction of (a) FM nanoclusters V(SO) and (b) CO nanoclusters 
V(CO) with and without an external magnetic field $h$ for $\lambda = 1.7$ and $\Delta$ = 0.05. 
Resistivity $\rho$ in units of ${\hbar a}/{\pi  e^2 }$ vs. $T$ with and without $h$ for 
(c) $\lambda = 1.7$ and $\Delta$ = 0.05, and (d) $\lambda = 1.5$ and $\Delta$ = 0.1. 
Filled symbols in (a)--(c) highlight that V(SO) and V(CO) remain unaffected, 
but the resistivity increases for $T$ = 0.05 and 0.055 with $h$.
}\label{fig:vco_h}
\end{figure}
%%%%%%%%%%%%%%%%%%%%%%%%%%%%%%%%%%%%%%%%%%%%%%%%%%%%%%%%%%%%%%%%%%%%%%%%

We now apply a very small external magnetic field $h$ to further confirm the phase coexistence 
above $T_{\rm c}$. For $h$ = 0.002, the $T_{\rm c}$ increases from 0.045 to 0.05 
[Figs.~\ref{fig:vco_h}(a) and (b)]. 
For $T\ge0.05$, V(CO) and V(SO) 
remain unaffected with $h$. This indicates that the applied magnetic 
field does not affect the CO regions, but only aligns spins in different FM nanoclusters 
without increasing their size, as shown schematically in Fig.~\ref{fig:schematic}(c). 
For $h$ = 0, magnetic nanoclusters are randomly oriented. Therefore, in addition to the CO 
nanoclusters, the up-spin (down-spin) electrons scatters from the down-spin (up-spin) oriented 
magnetic nanoclusters and the current is carried equally in both spin channels. A small 
external magnetic field aligns the magnetic nanoclusters, say in the up direction, which 
decreases the current from the down-spin electrons. This results in an overall enhancement 
of the resistivity, as shown in Fig.~\ref{fig:vco_h}(c) using filled symbols, and 
substantiates the presence of the FM nanoclusters above $T_{\rm c}$, in addition to the 
CO nanoclusters. Such enhancement of resistivity in a magnetic field due to the presence 
of FM and CO nanoclusters also resemble with FM-metallic/insulating multilayers in which 
spins in the individual FM layers are randomly oriented~\cite{pradhan3,sheng,zhou}. 
In addition, we find in Fig.~\ref{fig:vco_h}(d) that the resistivity for $\lambda = 1.5$ 
and $\Delta$ = 0.1 does not increases in a small external magnetic field ($h$ = 0.002) 
around $T_{\rm c}\,(=0.075)$ because of the absence of CO nanoclusters (discussed below). 

%%%%%%%%%%%%%%%%%%%%%%%%%%%%%%%%%%%%%%%%%%%%%%%%%%%%%%%%%%%%%%%%%%%%%%%%                               
\begin{figure} [t]
\centerline{
\includegraphics[width=8.75cm,height=3.50cm,clip=true]{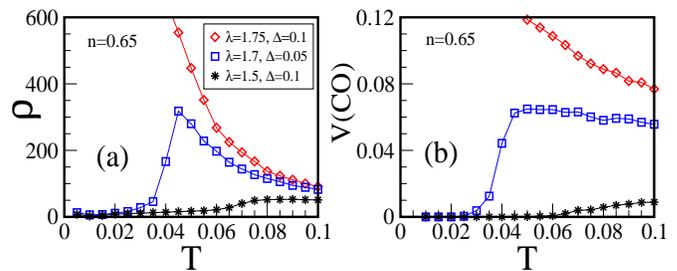}}
\caption{
$T$ dependence of (a) the resistivity $\rho$ in units of ${\hbar a}/{\pi  e^2 }$ and
(b) V(CO) for $\lambda$ = 1.75, 1.7, and 1.5. $\Delta$ values
are mentioned in figure. Legends in (a) and (b) are the same.
}
\label{fig:rho_vco}
\end{figure}
%%%%%%%%%%%%%%%%%%%%%%%%%%%%%%%%%%%%%%%%%%%%%%%%%%%%%%%%%%%%%%%%%%%%%%%%

Our overall results show that the phase competition between CO and FM nanoclusters above
$T_{\rm c}$ induces the resistivity hump, which is observed in LCMO. However, for large
bandwidth manganites, e.g. LSMO (for an abbreviation, see Table~\ref{table}), the
resistivity hump diminishes around $T_{\rm c}$ [see Fig.~\ref{fig:schematic}(b)]. In order
to understand this, we show the resistivity and V(CO) for $\lambda$ = 1.5, 1.7, and 1.75
in Fig.~\ref{fig:rho_vco}(a) and Fig.~\ref{fig:rho_vco}(b), respectively. Recall that
$\lambda$ is measured in
units of $t$, and thus smaller $\lambda$ corresponds to larger bandwidth or vice versa.
For $\lambda$ = 1.5 (1.75), the ground state is FM metallic (CE-type insulating) both at
$n$ = 0.65 and $n$ = 0.5, similar to LSMO (PCMO) [see Fig.~\ref{fig:schematic}(a)]. A
qualitative estimation of $\lambda$ for different manganites is listed in Table~\ref{table}.
For $\lambda$ = 1.5, V(CO) is very small and monotonically decreases with decreasing $T$,
resulting in no resistivity hump. This also suggests that the CO nanoclusters above $T_{\rm c}$
at $n$ = 0.65 appears only for specific systems for which the ground state at $n \sim 0.5$ is
a CE-type phase. LSMO does not have the CE-type phase at any hole doping unlike LCMO
[see Fig.~\ref{fig:schematic}(a)] and as a result the resistivity hump vanishes~\cite{coey}.
For $\lambda = 1.75$ i.e. small bandwidth manganites such as PCMO V(CO) increases with
decreasing $T$ and eventually an insulating state appears at low temperatures. We have
considered $\Delta$ = 0.1 for LSMO ($\lambda$ = 1.5) and PCMO ($\lambda$ = 1.75) as the
disorder strength is larger than LCMO ($\lambda$ = 1.7).

\section{Comparison with Experiments}

\begin{figure} [t]
\centerline{
\includegraphics[width=8.75cm,height=7.25cm,clip=true]{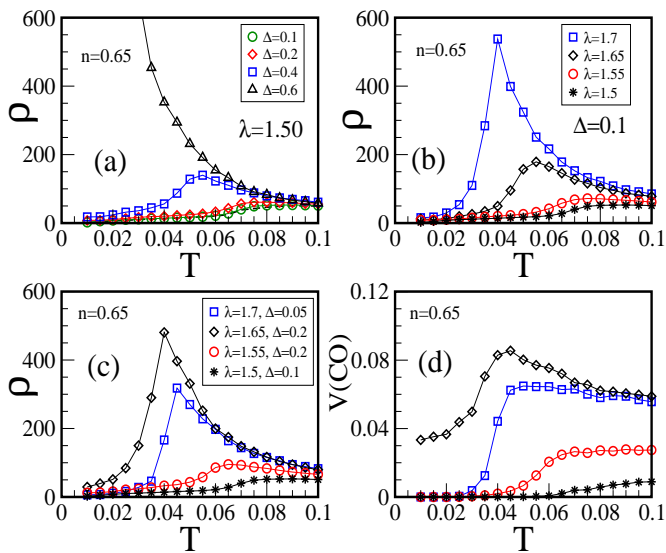}}
\caption{
(a)--(c): $T$ dependence of the resistivity $\rho$ in units of ${\hbar a}/{\pi  e^2 }$ for
different parameters (indicated in the figures). In (a), the dashed line is for $\Delta=0$.
(d): Temperature dependence of V(CO) for the same set of parameters used in (c).
Legends in (c) and (d) are the same.
}
\label{fig:rho}
\end{figure}

We turn now to explore the impact of disorder. As shown in Fig.~\ref{fig:rho}(a), the
resistivity hump appears and the magnitude increases with $\Delta$, even for $\lambda = 1.5$,
and its position shifts to lower temperature similar to the case for $\lambda = 1.7$
[Fig.~\ref{fig:pd}(d)]. However, relatively large $\Delta$ is required to convert the FM metal
to a disorder-assisted insulator for smaller $\lambda$.
Also, for $\Delta$ = 0.1, $T_{\rm c}$
increases but the magnitude of resistivity hump decreases with decreasing $\lambda$ (equivalent
to increasing the bandwidth) [see Fig.~\ref{fig:rho}(b)]. This is similar to the experiments
shown schematically in Fig.~\ref{fig:schematic}(b)~\cite{coey,saitoh,furukawa} except for NSMO.
Although LCMO has the smaller bandwidth than NSMO,
the magnitude of the resistivity hump for LCMO is smaller than NSMO.
In order to explain the observed resistivity trend properly, one needs
to estimate $\Delta$ values correctly, at least qualitatively. Due to large mismatch between R
and A radii, the disorder strength in NSMO is larger than LCMO and thus relatively
larger $\Delta$ must be set (see Table~\ref{table}).
As discussed earlier, $\Delta$ for LCMO is minimal,
while it increases for LSMO and increases further for NSMO and PSMO.
Figure~\ref{fig:rho}(c) shows the resistivity vs. temperature for four combinations of $\lambda$
and $\Delta$ values, corresponding to four different manganites (NSMO, LCMO, PSMO, and LSMO)
listed in Table~\ref{table}. Resistivity curves in Fig.~\ref{fig:rho}(c) qualitatively agree
with the experiments~\cite{saitoh,coey,furukawa} shown schematically in
Fig.~\ref{fig:schematic}(b). This shows that the disorder also plays an important role to
understand the experimental results systematically. V(CO) plotted in Fig.~\ref{fig:rho}(d)
shows a similar trend and clarifies the one-to-one correspondence between the resistivity
and the volume fraction the CO nanoclusters in CMR manganites.

%%%%%%%%%%%%%%%%%%%%%%%%%%%%%%%%%%%%%%%%%%%%%%%%%%%%%%%%%%%%%%%%%%%%%%%% 
%%%%%%%%%%%%%%%%%%%%%%%%%%%%%%%%%%%%%%%%%%%%%%%%%%%%%%%%%%%%%%%%%%%%%%%% 

\section{Conclusion}

In summary, based on the two-band model, we have provided a systematic study to demonstrate 
the phase competitions between the CO and the FM nanoclusters above $T_{\rm c}$ in CMR 
manganites. The resistivity increases due to the enhancement of the volume fraction of 
the $n$ = 1/2-type ($\pi, \pi$) CO and FM nanoclusters, with decreasing $T$ until 
$T_{\rm c}$. The FM nanoclusters start to grow and merge, and wins the competition below 
$T_{\rm c}$, leading to the sharp drop in the resistivity. The CO nanoclusters do not 
form in large bandwidth manganites and as a result the resistivity hump vanishes. 
Our calculations establish a simple yet complete pathway to understand the phase 
competitions in CMR manganites.

\section{Acknowledgements}

We acknowledge use of TCMP computer cluster at SINP and our discussions with P. Majumdar 
and S. K. Das. A part of the numerical calculations have been done using the computational 
resource on HOKUSAI GreatWave supercomputer allocated by RIKEN Advanced Center for 
Computing and Communication (ACCC).

% -----------------------------------------------------

% --------------------------------------------------------------
\end{document}